\documentstyle[preprint,pra,aps]{revtex}
\begin{document}
\draft
\title{Interesting thermomagnetic history effects in the 
antiferromagnetic state of SmMn$_2$Ge$_2$}
\author{S. B. Roy$^1$, S. Chaudhary$^1$, M. K. Chattopadhyay $^1$, P. Chaddah$^1$ and E. V. Sampathkumaran$^2$}
\address{$^1$Low Temperature Physics Laboratory,
Centre for Advanced Technology,Indore 452013, India\\
$^2$Tata Institute of Fundamental Research, Mumbai 400005, India}
\date{\today}
\maketitle
\begin{abstract}
We present results of magnetization measurements showing 
that the magnetic response of the antiferromagnetic state of 
SmMn$_2$Ge$_2$ depends on the path used in the 
field(H)-temperature(T) phase space to reach this state. Distinct 
signature of metastablity is observed in this 
antiferromagnetic state when obtained via field-cooling/field-warming paths.
The isothermal M-H loops show lack of end-point memory, reminiscent of that seen in metastable vortex states near the field-induced first order phase transition in various type-II superconductors.  
\end{abstract}                          
The intermetallic compound SmMn$_2$Ge$_2$ with its interesting 
magnetic properties has been a subject of intensive study during 
last two decades\cite{1,2,3,4,5,6,7,8,9,10}. In low applied 
magnetic fields it shows 
at least three magnetic transitions as a function of 
temperature\cite{1,2,3,6,7}. First it undergoes a paramagnetic (PM) to      
ferromagnetic (FM1) transition at around 350K, followed by 
an FM1 to antiferromagnetic (AFM) transition at around 160 
K (T$_{N1}$). On reduction of the temperature further this AFM state   
transforms again into another ferromagnetic (FM2) state around 100 
K(T$_{N2}$). There is a large spread in the reported values of the 
transition temperatures from FM1 to AFM and AFM to FM2 
states. Quality of the samples may be 
a possible source for the reported differences in the 
transition temperatures, especially when it is known that   
the microscopic magnetic properties of RE(rare-earth)1-2-2
compomuds are quite sensitive to their underlying crystal   
lattice structure. On the other hand, there exist now enough 
evidences from various studies that both of these transitions 
are probably first order in nature \cite{2,4,11,12}.  The first order nature 
of these magnetic phase transitions can also provide a 
natural explanation to the reported spread in the transition temperatures. 
Supercooling(superheating) can take place down(up) to a 
temperature T$^*$ (T$^{**}$) while cooling (heating) across
a first order transition point (T$_N$) \cite{13}. 
The extent of supercooled/superheated 
phases will depend on the path followed in the 
field(H)-temperature(T) phase space \cite{14}. In addition in 
the samples with defect structures the lower(higher) 
temperature phase will start nucleating around these defect 
structures once the sample is cooled(heated) across T$_N$.
This nucleation of the lower(higher) T phase will be 
completed  at T$^*$(T$^{**}$), and in the temperature 
regime T$_N$-T$^*$ (T$^{**}$-T$_N$) there will be co-existence of two
phases. All these properties will give rise to thermal 
hysteresis, and such thermal hysteresis is actually 
observed across the FM2-AFM and AFM-FM1 phase transitions
in SmMn$_2$Ge$_2$ \cite{2,4,7}.

Confined between two FM phase at low and high temperature 
and reached via first order phase transitions, the AFM 
phase in SmMn$_2$Ge$_2$ is something special. In this 
paper,  based on our careful dc        
magnetization studies we shall show that the magnetic 
response of this AFM state actually depends on the path used in
the (H,T) phase space to reach this state. Distinct 
trace of the high(low) temperature FM1(FM2) state remains well 
inside the AFM state, when this state is reached via a field 
cooling(warming) path. We seek a possible explanation of the
observed behaviour in terms of  supercooling/superheating 
and phase coexistence across a first order phase transition.

The SmMn$_2$Ge$_2$ sample used in our present study was     
prepared by argon arc melting and characterized by 
X-ray diffraction(XRD) measurements \cite{6,8}. Dc magnetization     
measurements were performed with a 
commercial SQUID-magnetometer (MPMS-5, Quantum Design)
using a 4 cm scan length.

A low field ( H=50 Oe) magnetization(M) versus temperature(T)
measurement reveal the FM1-AFM and AFM-FM2 transition      
temperatures in the present SmMn$_2$Ge$_2$ sample to be 
approximately 150 K and 105 K respectively (see Fig.1). We thus 
choose a temperature T=120 K which is well inside the AFM 
state away from both the upper and lower temperature
phase boundaries. Moreover, the M-H measurements at 
120K reveal a field-induced 
ferromagnetic transition around H$\approx$4 kOe 
(see the inset of Fig.1). 
To keep the sample well inside the AFM state in the
(H,T) space,  the applied field in the 
present measurements is limited to a maximum 
value of H$_{max}$=1 kOe . We present in Fig.2(a) the M-H 
plot at T=120 K, measured after reaching this temperature    
in zero field cooled (ZFC) condition. There is a small non-linearity
as well as a small but 
distinct hysteresis in this M-H curve, which indicates that the 
magnetic state is not pure antiferromagnetic in nature and 
there probably exists a finite amount of spin-canting. The same feature may
also arise from very small amount (1-2\%) of ferromagnetic phase which can
go undetected in  standard XRD measurements\cite{6,8}. In fact similar 
non-linearity in the M-H curves in the antiferromagnetic state of various 
CeFe$_2$ based pseudobinary alloys, has earlier been attributed to small amount of ferromagnetic impurity phase which could not be detected in the
XRD studies\cite{15}. However, subsequent studies have indicated that this feature can be of intrinsic origin\cite{16,17,18}. In our present system also
various studies (to be narrated below) indicate that the unconventional properties of the antiferromagnetic state in SmMn$_2$Ge$_2$ cannot be explained in terms of a small amount of ferromagnetic impurity phase.

We shall now study the magnetic response of the AFM state at 
120K after reaching this state following the three 
different experimental protocols :
\begin{enumerate}
\item the temperature 120 K is reached from temperature well 
above T$_{N1}$ in absence of any applied field, and then a 
field of 1 kOe is switched on.  
\item a field of 1 kOe is switched on within the FM2 state 
at 4.5K and the AFM state is reached subsequently by 
warming up the sample unidirectionally across T$_{N2}$ to 120K. 
\item a field of 1 kOe is switched on within the FM1 state 
at 300 K  and the AFM state is reached subsequently by cooling the sample 
unidirectionally across T$_{N1}$ to 120K.
\end{enumerate}

In all these experimental protocols sufficient wait time is given after reaching 120K to ensure complete temperature stability. Furthermore, in experimental protocol no. 1 and 3 temperature is reduced slowly in small steps to avoid any temperature oscillation. 

The values of magnetization measured at 120 K with 
H=1 kOe in both the protocols 2 and 3 
are higher than that measured after switching on the 
field of 1 kOe at 120K in the ZFC condition i.e.with the protocol 1. 
This observation can be rationalized in terms 
of supercooling (superheating) of the 
FM1(FM2) state. While cooling(warming) acrosss the 
FM1(FM2)-AFM transition temperatures 
T$_{N1}$(T$_{N2}$) some amount of the 
FM1(FM2) state will supercool (superheat) into the 
temperature regime well beyond the transition temperature. 
The extent of temperature regime    
of supercooling/superheating actually widens in presence of  
the applied magnetic field \cite{18,19}. 

To support the above conjecture we have studied the 
detailed nature of the magnetic state at T=120K obtained by three 
different experimental protocols mentioned above. We use
a "$\it{minor}$ $\it{hysteresis}$ $\it{loop}$ $\it{technique}$" to show that
the AFM state at 120K obtained in ZFC 
condition is a stable magnetic state, while the magnetic 
states obtained at 120K by cooling/heating in presence of 
1 kOe field are metastable in nature. 
Taking the M-H curve drawn in the ZFC AFM state at 120K
by field cycling between 1 kOe and 0 kOe as the envelope curve, 
we draw minor hysteresis loops (MHLs), terminating 
the M-H curve at various field points ($0<H<1 kOe$) on 
the the descending field leg of the envelope curve (see Fig. 2(a) and 3(a)). 
A distinct "$\it{end}$ $\it{point}$ $\it{memory}$" is observed for 
all the MHls, namely on completion 
of the cycle the MHLs show the same end point magnetization value    
on the envelope M-H curve at 1 kOe. 
This kind of "$\it{end}$ $\it{point}$ $\it{memory}$" is common with various 
kinds of hysteretic systems including hard ferromagnets and 
superconductors \cite{20,21}. To emphasise this point 
we show in  Fig. 2 (b) and 2 (c) similar M-H curves obtained 
by field cycling between
1 k Oe and 0 Oe at T=20K and 220 K which are well inside the 
FM2 and FM1 phase respectively . Clear 
"$\it{end}$ $\it{point}$ $\it{memory}$"  is observed in both the cases. 

We shall now draw the envelope curve and the MHLs in the 
AFM state at 120K obtained by heating from 4.5 K in 
the presence of a field of 1 kOe i.e. following the protocol
no.2. In contrast to the ZFC 
case, the envelope curve and the MHLs are of very different    
nature (see Fig.3(b)).The most prominent difference is the 
absence of "$\it{end}$ $\it{point}$ $\it{memory}$". Starting from the 
H=1 kOe point and returning back
to this point by drawing MHLs of increasing field amplitude,
a steadily decreasing value of magnetization is observed at the
end point i.e. H=1 kOe. The end point memory is in fact lost
in the process of drawing the first MHL by cycling between 
1 kOe and 0.8 kOe (see the paths marked by 1 and 2 
in the inset of fig. 3(b)).
The envelope curve drawn by lowering the field from 1 kOe after this
first cycle (see path marked by 3 in the inset of Fig.3) is
quite different from the initial envelope curve. The differences 
in both the end point magnetization and the envelope curves rise steadily 
with further cycling of field with succesively larger field
amplitude (see path marked by 5,6,7,8,9 and 10 in the
inset of Fig.3(b)).  Same field cycling process does not cause
any effect on the end point magnetization and envelope curve in the
magnetic state obtained with the experimental protocol no.1 (see Fig. 3(a)). This clearly shows that the initial magnetic
state obtained with protocol no. 2 is a metastable state, 
and the energy fluctuation     
introduced while drawing the MHLs steadily push this state
towards the stable ZFC state. This kind of the lack of 
"$\it{end}$ $\it{point}$ $\it{memory}$" has earlier been observed across the 
vortex matter phase transition from one kind of vortex solid to another
in a type-II superconductor CeRu$_2$ (see Fig. 7 of Ref.22 ). This was 
attributed to the existence of metastable states across a disorder 
broadened first order transition \cite{22}. These metastable states were
shattered through the energy fluctuation generated while drawing the MHLs
in the concerned (H,T) regime. The same behaviour has been observed 
subsequently across a disorder induced first order transition in the vortex
matter phase space of another type-II superconductor NbSe$_2$ \cite{23}.

The same  lack of "$\it{end}$ $\it{point}$ $\it{memory}$" effect is observed 
at 120 K on cooling form 300 K well 
inside the FM1 state i.e. following the protocol no.3 (see Fig. 3(c)). 
On the other hand same kind of experiments after preparing the magnetic states
well inside the ferromagnetic regions FM1 and FM2 by crossing the transition temperatures T$_{N1}$ and T$_{N2}$ in presence
of field, show clear "$\it{end}$ $\it{point}$ $\it{memory}$" effect
(data not shown here for the sake of clarity and conciseness). These results also negate any possible contribution from an impurity ferromagnetic phase in the observed metastable behaviour of the AFM state. Metastability (if any) related to the hindrance of domain motions in a ferromagnet is known to be more while warming up from the low temperature ZFC state than while cooling down from the high temperature region in the presence of an applied field\cite{24}.

This lack of "$\it{end}$ $\it{point}$ $\it{memory}$"
is now accepted as a signature  of metastability associated with a first order transition \cite{22,23}, which is taken as a support for establishing the first order nature of certain vortex matter phase transitions in type-II superconductors \cite{22,23,25}. In this paper we have looked at the AFM state of SmMn$_2$Ge$_2$ which can be reached from both high and low temperature FM state through magnetic phase transitions. The first order nature of these two transitions is already considered \cite{2,4,11,12}. We have now shown the lack of "$\it{end}$ $\it{point}$ $\it{memory}$" in the minor hysteresis loops and associated metastable behaviour in this AFM state of  SmMn$_2$Ge$_2$. The observed behaviour highlights the interesting status of the AFM state in SmMn$_2$Ge$_2$ sandwiched between 
two FM states in the (H,T) phase space and reached via first order phase
transitions. 
  
Summarising our results, we find interesting thermomagnetic 
history effects well inside the AFM state of 
SmMn$_2$Ge$_2$. When this AFM state is reached either from 
the FM1(FM2) state by cooling (heating) in presence of an 
applied field, the traces of the FM1(FM2) state remain
as supercooled(superheated) state. This is an example of 
phase coexistence of the AFM-FM1(FM2) state, and the resulting      
magnetic state is metastable in nature. On drawing MHLs in  
this metastable state one introduces energy fluctuations, 
which drive the domains of metastable FM1(FM2) state to the 
stable AFM state. Such metastability  
in the form of the lack of  "$\it{end}$ $\it{point}$ $\it{memory}$"
(which is also
observed across solid to solid vortex matter phase transitions
in various type-II superconductors (ref.22 and 23)) may 
serve as characteristic signatures of a first order 
transition in  samples with substantial defect structures 
where the detection of latent heat as the canonical       
signature of a first order transition is relatively 
difficult \cite{26}.

\begin{figure}
\caption{M vs T plot of SmMn$_2$Ge$_2$ in 
an applied field of 50 Oe. The sample was zero field cooled to the lowest
temperature before switching on the field, and data was taken while warming
up the sample. The inset shows M vs H curve drawn at T=120K after reaching
that temperature in zero field cooled condition.}
\end{figure}
\begin{figure}
\caption{M vs H plot of SmMn$_2$Ge$_2$ obtained by field cycling between 
0 and 1 kOe starting from zero field cooled condition 
at (a) 120 K, (b) 20 K and (c) 220K. The field cycling sequence
is the following: (1) the field is increased from 0 to 1 kOe, (2) decreased
from 1 kOe to 0 Oe (3) increased again from 0 to 1 kOe.}
\end{figure}
\begin{figure}
\caption{M vs H plot and minor hysteresis loops with field cycling between
0 and 1 kOe
of SmMn$_2$Ge$_2$ at 120 K obtained with three 
different experimental protocols (see text for details). 
(a) Results obtained with protocol no. 1.
(b) Results obtained with protocol no. 2. The inset shows the 
minor hysteresis loops drawn in the following field sequence. H is decreased
from 1 kOe to 0.8 kOe (path 1), increased from 0.8 Oe back to 1 kOe (path 2),
decreased from 1 kOe to 0.6 kOe (path 3), increased from 0.6 kOe to 1 kOe
(path 4), decreased from 1 kOe to 0.4 kOe (path 5), increased from 0.4 kOe to
1 kOe (path 6), decreased from 1 kOe to 0.2 kOe (path 7), increased from 0.2 kOe
to 1 kOe (path 8), decreased from 1 kOe to 0 Oe (path 9) and lastly increased
from 0 Oe to 1 kOe(path 10). The last two sequence essentially forms the
envelope curve. Note that the end point magnetization at H=1 kOe decreases 
progressively. The same set of sequence is followed for drawing minor hysteresis
loops fig. 3(a). In contrast to the present case the end point magnetization
at H=1 kOe retains the same value confirming end point memory.
(c) Results obtained with protocol no. 3. The inset shows the 
minor hysteresis loops drawn in the following field sequence. H is decreased
from 1 kOe to 0.8 kOe (path 1), increased from 0.8 Oe back to 1 kOe (path 2),
decreased from 1 kOe to 0.6 kOe (path 3), increased from 0.6 kOe to 1 kOe
(path 4), decreased from 1 kOe to 0.4 kOe (path 5), increased from 0.4 kOe to
1 kOe (path 6), decreased from 1 kOe to 0.2 kOe (path 7), increased from 0.2 kOe
to 1 kOe (path 8), decreased from 1 kOe to 0 Oe (path 9) and lastly increased
from 0 Oe to 1 kOe(path 10). The last two sequence essentially forms the
envelope curve. Note that the end point magnetization at H=1 kOe decreases 
progressively as in fig. 3(b), showing lack of end point memory.}    
\end{figure}
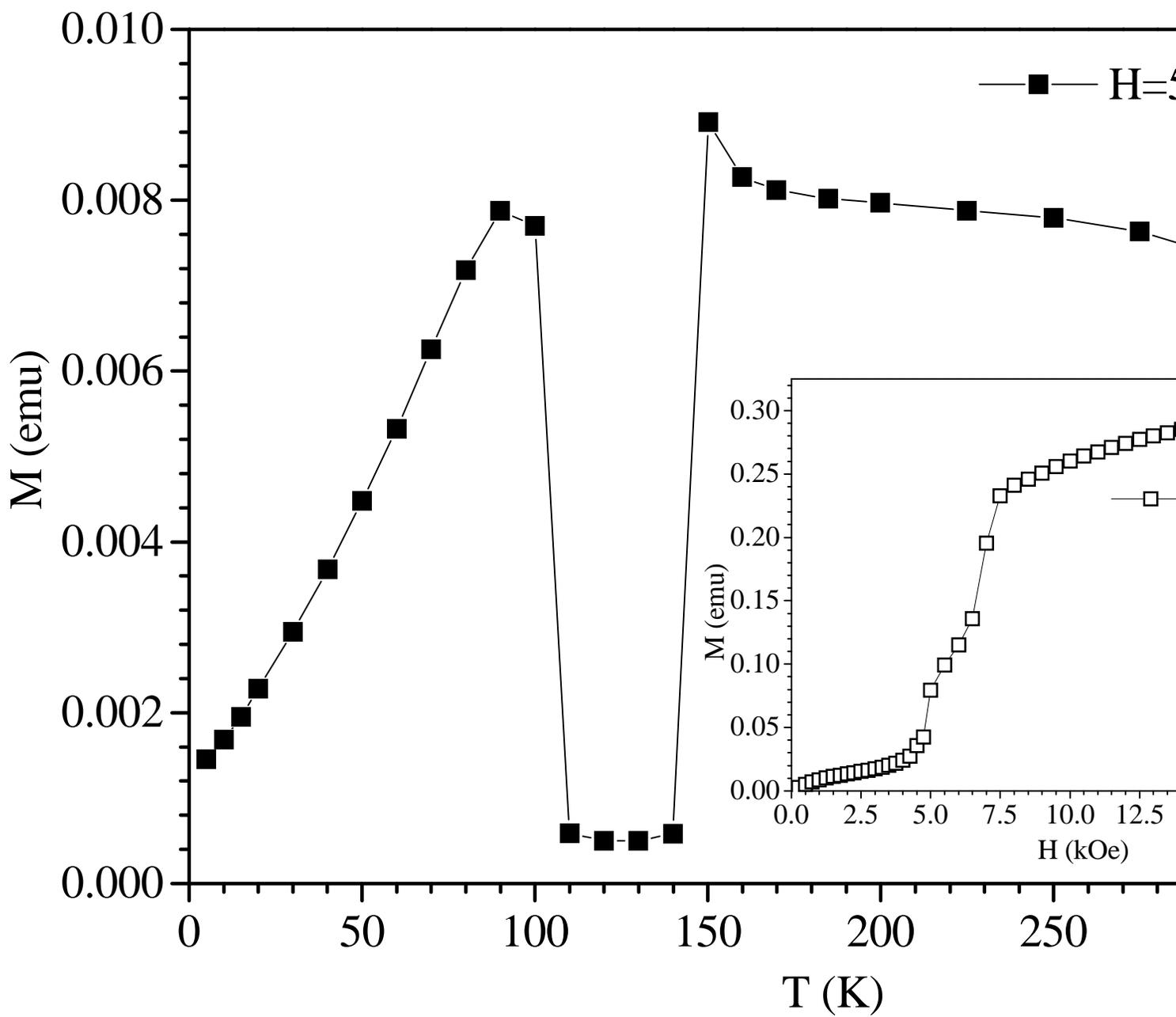
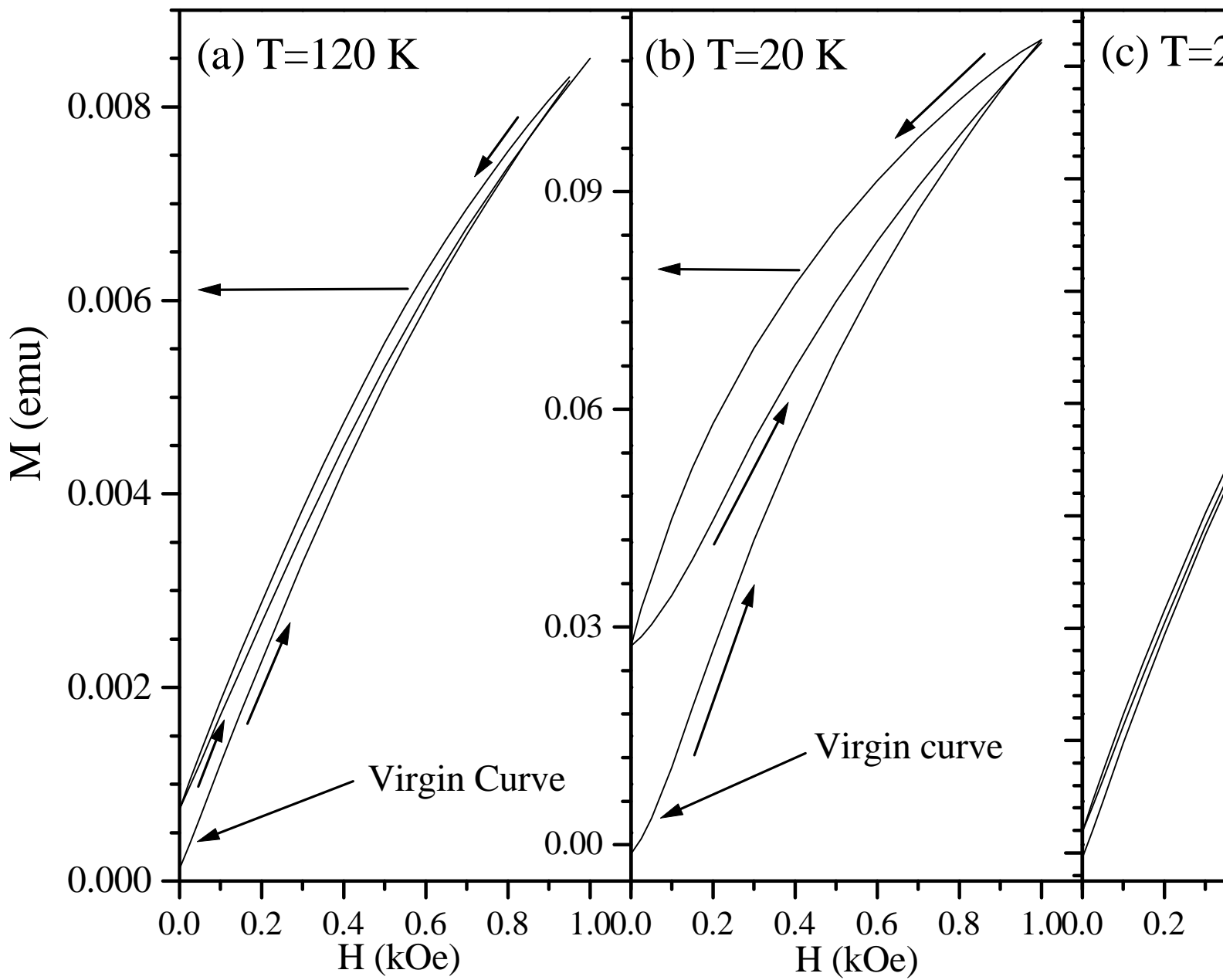
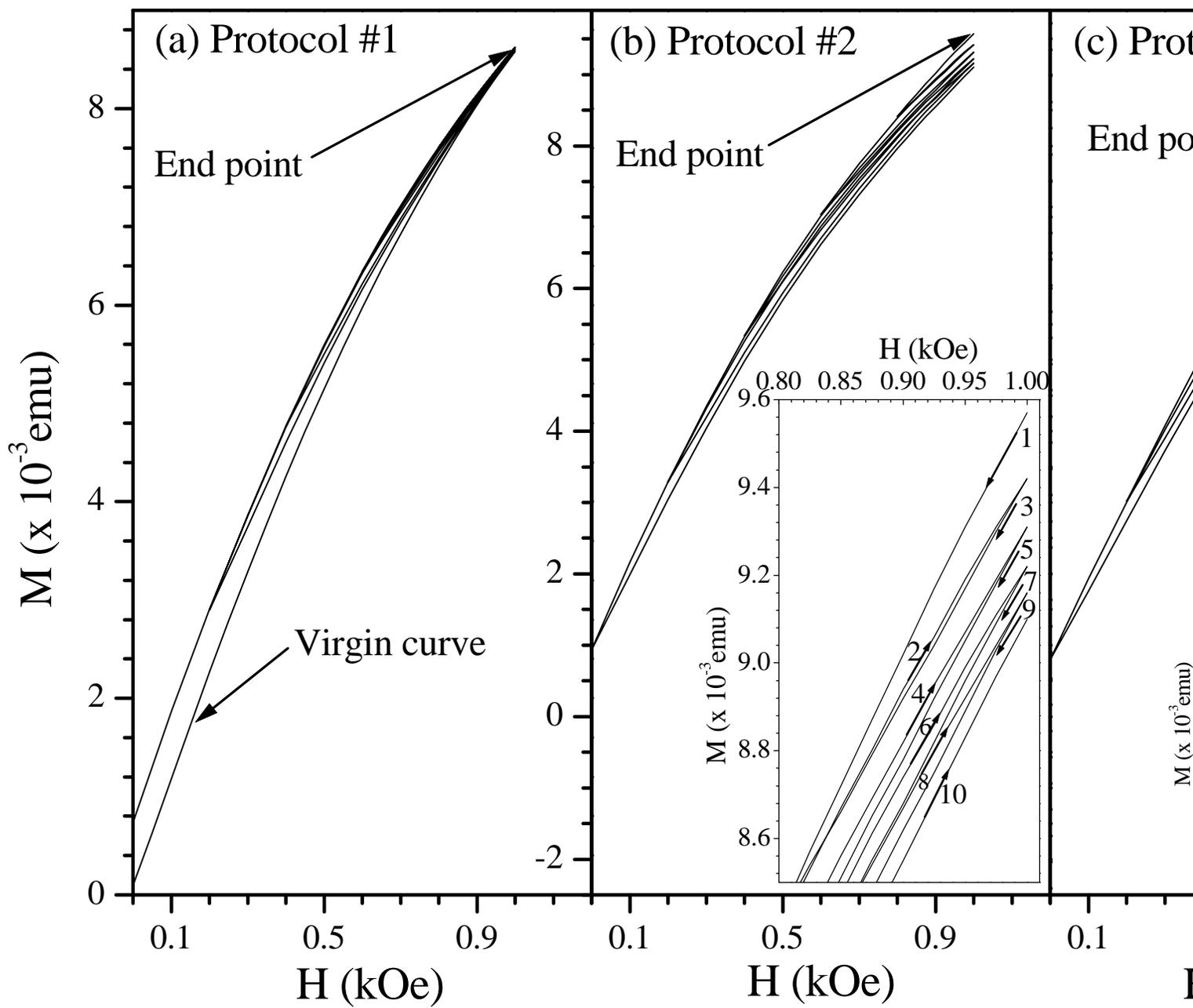
\end{document}